\documentclass[prl,twocolumn,showpacs]{revtex4}
\usepackage{graphics}
\usepackage{dcolumn}
\begin{document}
\title{Enhancing T$_c$ in field-doped Fullerenes by applying uniaxial stress}
\author{Erik Koch}
\email{E.Koch@fkf.mpg.de}
\affiliation{Max-Planck Institut f\"ur Festk\"orperforschung,
             Heisenbergstra\ss e 1, 70569 Stuttgart, Germany}
\date{\today}
\begin{abstract}
Capitalizing on the two-dimensional nature of superconductivity in
field-effect doped C$_{60}$, we show that it should be possible to
increase the transition temperature $T_c$ by applying uniaxial stress
perpendicular to the gate electrode. This method not only holds the
promise of substantially enhancing $T_c$ (by about 30 K per GPa),
but also provides a sensitive check of the current understanding 
of superconductivity in the doped Fullerenes. 
\end{abstract}
\maketitle

In a remarkable series of experiments Sch\"on and collaborators have
demonstrated superconductivity in field-effect devices based on Fullerene
crystals. In a first step they field-doped pure C$_{60}$ with
electrons, observing a maximum $T_c$ of 11 K at a doping of about 3 electrons
per molecule in the layer under the gate electrode \cite{elecdoped}.
Reversing the polarity of the gate voltage, they achieved hole-doping, finding
superconductivity with a transition temperature of up to 52 K at a doping level
of about 3 to 3.5 holes per molecule \cite{holedoped}. In the most recent
experiments, they observed superconductivity by field-doping C$_{60}$ crystals
that were intercalated with chloroform, CHCl$_3$, and CHBr$_3$
\cite{latticeexp}. In these crystals, the volume per C$_{60}$ molecule is
increased due to the presence of the intercalated molecules. Superconductivity
was observed for electron- as well as for hole-doping and it was found that in
both cases $T_c$ increases almost linearly with the distance between
neighboring C$_{60}$ molecules, reaching 80 K for C$_{60}\cdot$2CHCl$_3$ 
and 117 K for C$_{60}\cdot$2CHBr$_3$. After this discovery a 'race to beat
the cuprates' has been announced \cite{dagotto}, and the search is on
for ways to further increase the transition temperature in the Fullerenes.
Here we show that a feasible method for doing so is the application of
uniaxial stress perpendicular to the gate electrode. While this proposal
might at first seem counterintuitive, we will demonstrate that it is a natural
consequence of the two-dimensional nature of superconductivity in the 
field-doped Fullerenes.

So far superconductivity in field-doped Fullerenes has been discussed
in close analogy to alkali-doped C$_{60}$. This was motivated by the 
remarkable similarities between these two classes of materials \cite{gunnar}:
(i) in both $T_c$ is largest when the $t_{1u}$ band is half-filled, and
(ii) $T_c$ increases with the distance between the C$_{60}$ molecules in
the crystal, which is explained as a consequence of the corresponding increase
in the density of states at the Fermi level. Experimentally it is found that
for a given nearest-neighbor distance the transition temperature for the 
alkali- and the electron-field-doped Fullerenes are almost identical 
\cite{latticeexp}.
Nevertheless, there is a fundamental difference between the two classes of
materials: the alkali-doped Fullerenes are bulk-superconductors, while in 
the field-effect devices superconductivity is restricted to two dimensions. 
It appears that in the field-doped Fullerenes only the first monolayer
under the gate electrode is doped. This is confirmed by tight-binding 
simulations where it is found that that the deeper layers carry
negligible charge density \cite{wehrli}. Experimentally it is supported
by the fact that for the increase in $T_c$ upon intercalation of CHCl$_3$ 
and CHBr$_3$ only the change in the nearest neighbor distance seems to matter, 
while the change in crystal structure appears to have no effect: At the 
temperatures where superconductivity occurs, pure C$_{60}$ is simple cubic 
with four molecules (sitting on fcc sites) per unit cell \cite{david}, while 
the intercalated crystals are (almost) hexagonal \cite{jansen,nodos}. This 
suggests that superconductivity in field-doped Fullerenes is restricted to a 
single, triangular layer: a (111) plane in C$_{60}$ and a (001) plane in 
C$_{60}\cdot$2CHCl$_3$ and C$_{60}\cdot$2CHBr$_3$.

Given that only a single lattice plane is involved in the superconductivity
and given that the transition temperature $T_c$ increases with the distance 
between neighboring molecules, it should be possible to increase $T_c$ simply
by pushing on the gate electrode of the field-effect device. Such uniaxial
stress will decrease the spacing of the lattice planes parallel to the gate,
but at the same time it will {\em increase} the distance of the molecules  
{\it in} the planes. This is the Poisson effect \cite{nye}.
We again stress the fundamental difference between the alkali- and the 
field-doped Fullerenes. The alkali-doped materials are bulk superconductors.
Thus under pressure, even under uniaxial stress, the volume per molecule is
reduced, lowering the density of states at the Fermi level and hence $T_c$.
In contrast, in the field-doped Fullerenes only the layer under the gate
electrode is doped, and therefore hopping to the deeper layers is suppressed 
by the strong electrostatic potential of the induced space charge \cite{wehrli}.
Thus, reducing the distance between the layers will have a negligible effect
on the (practically two-dimensional) density of states, which is, however, 
strongly increased by the increasing distance of the molecules in the plane.

\begin{figure}
 \resizebox{3in}{!}{\includegraphics{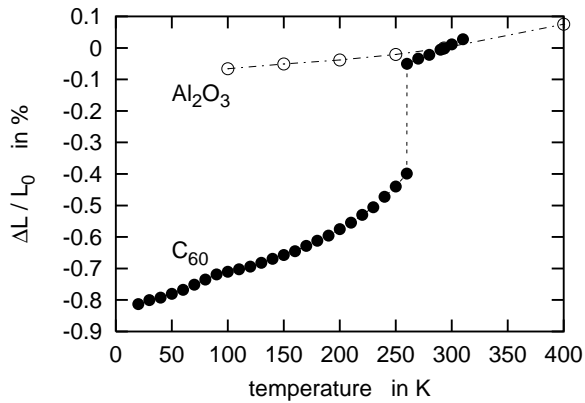}}
 \caption[]{\label{thermplot}
  Linear thermal expansion of a C$_{60}$ single crystal \cite{david} and
  of Al$_2$O$_3$ \cite{thermexp}.}
\end{figure}

At this point the obvious questions are: Will uniaxial stress really
affect the molecules under the gate electrode? And if so, how large will 
the effect be? 
The first problem that comes to mind is that the molecules under the gate
might be held in place through interactions with the Al$_2$O$_3$ gate oxide. 
In this case the response of the molecules to stress would rather be determined 
by the hard oxide than by the elastic constants of the soft organic crystal.
Strong bonding of the C$_{60}$ molecules to the gate dielectric seems, however,
quite unlikely, and there is even direct evidence for the mobility of the
molecules that carry the source-drain current: Upon cooling, solid C$_{60}$
undergoes an orientational phase transition at about 260 K, which is
accompanied by a substantial reduction in volume (the distance between the
molecules decreases by more than 0.3\%, see Fig.~\ref{thermplot}) \cite{david},
while the thermal expansion of aluminum oxide does not show any significant 
structure in that temperature range \cite{thermexp}. Nevertheless, the ordering
transition shows up in the resistivity of field-doped C$_{60}$ (Fig.~5 of
Ref.~\cite{holedoped}), indicating that the molecules in the relevant
layer indeed remain mobile.

The next question is whether it is experimentally feasible to apply pressure
of any significant magnitude to field-effect devices based on delicate organic
crystals.  As it turns out, the necessary techniques are well
established for the investigation of quasi-two-dimensional organic conductors:
The whole device is embedded in a suitable epoxy or frozen in oil, and the
application of pressures of up to 1 GPa at temperatures down to 0.5 K to
samples of millimeter dimensions is straightforward \cite{pressure1,pressure2}.
Ironically, one of the major problems in such experiments seems to be the
suppression of the Poisson effect \cite{pressure2}, the very effect which is
desired in the present case.
Thinking about C$_{60}$ under pressure one might worry about polymerization.
Pressures of the order of 1 GPa can, however, be considered low
\cite{sundqvist}. Moreover, under uniaxial stress the crystal will rather 
loose cohesion and be squeezed, than that the resilient C$_{60}$ molecules 
will form chemical bonds.

How large an effect on $T_c$ can we expect? For an estimate we calculate
the Young modulus $E$ and the Poisson ratio $\sigma$ for C$_{60}$ under
uniaxial stress. 
In terms of the elastic constants, the elastic response to stress in the 
[100] direction is given by
\begin{displaymath} 
 E_{[100]}={(c_{11}-c_{12})(c_{11}+2c_{12})\over c_{11}+c_{12}}
 \;,\quad
 \sigma_{[100]}={c_{12}\over c_{11}+c_{12}} .
\end{displaymath}
For stress in the [111] direction we find 
\begin{displaymath}
 E_{[111]}={3(c_{11}+2c_{12})c_{44}\over c_{11}+2c_{12}+c_{44}}
 \;,\quad\!
 \sigma_{[111]}={c_{11}+2(c_{12}-c_{44})\over 2(c_{11}+2c_{12}+c_{44})} .
\end{displaymath}

\begin{table}
 \begin{center}
 \begin{tabular}{ccc|c|cc|cc|l}
  $c_{11}$&$c_{12}$&$c_{44}$ & $B$ & $E_{[100]}$&$\sigma_{[100]}$ & 
                                     $E_{[111]}$&$\sigma_{[111]}$\\\hline
  14.1& 6.0& 7.7 &  8.7 & 10.5&0.30 & 17.8&0.16 & \cite{yildharris} pot.~II\\
  14.9& 6.9& 8.1 &  9.6 & 10.5&0.32 & 19.0&0.17 & \cite{yildharris} pot.~I\\
  16.1& 8.2& 8.2 & 10.8 & 10.5&0.34 & 19.6&0.20 & PR potential\\
  24.4&12.4&12.4 & 16.4 & 16.1&0.34 & 29.7&0.20 & GF potential\\
  24.5&12.3&11.8 & 16.4 & 16.3&0.33 & 28.5&0.21 & \cite{burgos}
 \end{tabular}
 \end{center}
 \vspace{-2ex}
 \caption[]{\label{elastconst}
  Elastic constants for a C$_{60}$ crystal at $T=0$. 
  Elastic constants, bulk modulus ($B=(c_{11}+2c_{12})/3$), and
  Young moduli $E$ are in GPa; Poisson ratios $\sigma$ are dimensionless.}
\end{table}

Table \ref{elastconst} gives a list of representative theoretical values 
for the elastic constants of C$_{60}$ from the literature and from calculations
using the Girifalco \cite{girifalco} and the Pacheco-Ramalho \cite{pacheco}
potentials. There is a remarkably large spread in the predicted elastic 
constants, which vary by up to a factor of two. The Young moduli fall in the
ranges $E_{[100]}\approx10\ldots16$ GPa and $E_{[111]}\approx18\ldots30$ GPa.
The Poisson ratios differ, however, by only about ten percent:
$\sigma_{[100]}\approx0.32$ and $\sigma_{[111]}\approx 0.2$.
Using the slope $c\approx$ 230 K/\AA\ of $T_c$ as a function of $a=\sqrt{2}\,d$,
where $d$ is the distance to the nearest molecule in the doped layer, that
was found experimentally \cite{latticeexp}, we can estimate the increase in
transition temperature under uniaxial stress: $\Delta T_c/p=ca_0\sigma/E$.
For a hole-doped C$_{60}$ crystal under uniaxial stress in the [111] direction
we thus find $\Delta T_c^{[111]}/p\approx 22\ldots 36$ K/GPa. Should the [100] 
plane be the relevant for superconductivity, the effect would be even larger:
$\Delta T_c^{[100]}/p\approx 65\ldots105$ K/GPa.

\begin{figure}
 \centerline{\resizebox{3in}{!}{\includegraphics{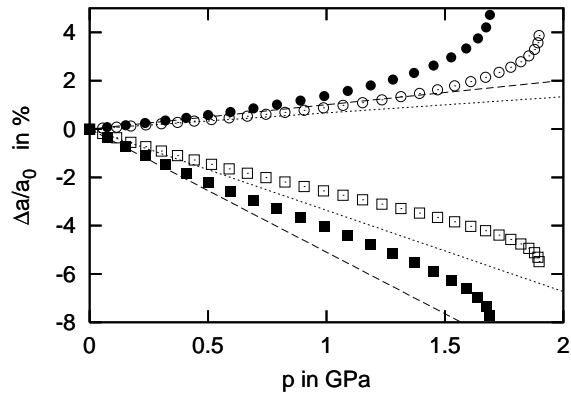}}}
 \caption[]{\label{fig111}
  Response to uniaxial stress in [111] direction
  calculated using the Girifalco (open symbols and dotted lines) and
  the Pacheco-Ramalho (filled symbols and dashed lines) potential.
  The lines show the deformation expected from the Young modulus,
  the symbols show the calculated deformations under finite stress.}
\end{figure}

The effect of uniaxial stress on the transition temperature in field-doped
Fullerenes should thus be large, indeed. Already a modest force of 
20 Newton on a crystal with an area of a square millimeter should increase
$T_c$ by about half a Kelvin or more. Such an effect should be observable
when monitoring the source-drain resistivity of the field-effect device
just above the ambient-pressure transition temperature.  
The increase in $T_c$ is of course limited by the yield stress of the
C$_{60}$ crystal. To get a feeling for the behavior of C$_{60}$ under
finite stress, Fig.~\ref{fig111} shows the increase in the intermolecular
distance $d$ in the plane and the decrease in distance between the lattice 
planes as a function of the applied stress in [111] direction. For finite 
stresses the increase in $d$ is even larger than expected from the elastic
constants. Eventually the slope of $\Delta d/d_0$ becomes, however, infinite, 
which means that the crystal is squeezed.

In our estimate of the change in transition temperature under uniaxial stress
we have assumed that the interpretation given in Ref.\ \onlinecite{latticeexp}
is correct, namely that the distance of the molecules in the doped lattice 
plane is the only relevant parameter for determining $T_c$. A recent structural
analysis of the low temperature phases of C$_{60}\cdot$2CHCl$_3$ and 
C$_{60}\cdot$2CHBr$_3$ has, however, cast doubts on this interpretation
\cite{nodos}. It might therefore be possible, that other effects, like
the different molecular orientations or the presence of the intercalants
substantially influence superconductivity. In that respect, investigating
the change in $T_c$ under uniaxial stress could help to clarify the situation.

In summary, the application of uniaxial stress is a straightforward and 
feasible way for enhancing the transition temperatures of field-doped 
Fullerenes even further. We have shown that the enhancement is significant,
about 30 K per GPa, so that already the effect of very small pressures should 
be measurable. Besides the potential for achieving record $T_c$'s, such 
experiments would deepen our understanding of the physics behind the increase 
in transition temperature, since they would allow to study the effect of an 
increased spacing between the molecules without having to introduce additional 
molecules. 
Due to the softness of the Fullerene crystals it is tempting to speculate that
it might even be possible to push the field-doped Fullerenes across the 
Mott transition \cite{c60mott}. Then, by continuously varying the distance 
between the molecules (by the applied stress) and the doping (by the gate 
voltage) it should be possible to study the physics of a doped Mott insulator.
From the alkali-doped Fullerenes we know that $T_c$ decreases when the 
lattice constant is increased too much \cite{screen,TcMott}. It will thus be 
interesting to see whether $T_c$ in the field-doped Fullerenes will be limited 
by the Mott transition, or if the maximum $T_c$ is realized in the Mott 
insulating regime --- at some optimal doping.

It is a privilege to thank O.~Gunnarsson for sharing his insights and
for his continuous support.
We would also like to thank R.K.~Kremer, W.~Branz, J.~Merino, K.~Syassen, 
and J.H.~Sch\"on for helpful discussions.

\end{document}